\documentclass[12pt]{iopart}

\usepackage{amssymb}
\usepackage{amsbsy}
\usepackage{graphicx}
\usepackage{setstack} 
\usepackage{cite}  

\eqnobysec

\usepackage{epsfig}

\begin{document}

\title[Current distribution in magnetically confined 2DEG]{Current distribution in magnetically confined 2DEG: 
	semiclassical and quantum mechanical treatment
}

\author{R\'obert N\'emeth$^1$, Zolt\'an Kaufmann$^1$ 
	   and J\'ozsef Cserti$^1$}

\address{$^1$ 	ELTE E{\"o}tv{\"o}s Lor\'and University,
	Department of Physics of Complex Systems,
	H-1117 Budapest, P\'azm\'any P{\'e}ter s{\'e}t\'any 1/A, Hungary
	}

\begin{abstract}

In the ballistic regime we study both semiclassically and quantum mechanically the electron's dynamics in two-dimensional electron gas (2DEG) in the presence of an inhomogeneous magnetic field applied perpendicular to the plane. The magnetic field is constant inside four separate circular regions which are located at the four corners of a square of side length larger than the diameter of the circles, while outside the circles the magnetic field is zero. We carry out the stability analysis of the periodic orbits and for given initial conditions numerically calculate the two-dimensional invariant torus embedded in the four-dimensional phase space.
Applying the Bohr--Sommerfeld and the Einstein--Brillouin--Keller semiclassical quantization methods we obtain the energy levels for different magnetic field strengths. We also perform exact quantum calculations solving numerically the discretized version of the Schr\"odinger equation. In our calculations, we consider only those bound states that are localized to the neighborhood of the four magnetic disks. We show that the semiclassical results are in good agreement with those found from our quantum calculations.
Moreover, the current distribution and the phase of the different wave functions enable us to deduce the two quantum numbers $n_1$ and $n_2$ characterizing the energy levels in the semiclassical methods. Finally, we present two examples in which the quantum state shows a similar structure to the previous states, but these are special in the following sense. One of them is a scar state localized to the neighborhood of the periodic orbit while this orbit is already unstable. In the case of the other state, the current density is circulating in two rings in opposite direction. Thus, it is not consistent with the classical motion in the neighborhood of the periodic orbit.

\end{abstract}


\submitto{NJP}

\maketitle

\section{Introduction}
\label{intro:sec}

During the last few decades, significant progress can be observed in nanotechnology both in the experimental and theoretical sides. 
Using the well-developed platform the two-dimensional gas (2DEG) 
in high mobility heterostructures can easily be confined electrostatically~\cite{Carlo-Houten,Carlo-konyv,PhysRevLett.69.506,PhysRevLett.66.2790}.  
To confine the 2DEG another possible experimental method is to use spatially inhomogeneous magnetic fields. 
Such an inhomogeneity of the magnetic field can be realized experimentally, e.g., by varying the topography of the electron gas~\cite{Foden_1994,PhysRevB.52.R8629}, or using ferromagnetic materials~\cite{Leadbeater-ferro,Krishnan-ferro,micromagnet_dysprosium_114520}, or depositing a superconductor on top of the 2DEG~\cite{PhysRevB.50.14726,Geim1997}.
The experimental realization of such magnetic field generates considerable attention 
to study theoretically the two-dimensional dynamics of electrons propagating under the presence of a perpendicular inhomogeneous magnetic field~\cite{PhysRevLett.68.385,PhysRevB.47.1466,PhysRevB.48.2365,PhysRevB.48.15166,PhysRevLett.72.1518,PhysRevB.52.17321,PhysRevLett.80.1501,PhysRevB.60.8767,PhysRevB.63.235317,PhysRevB.64.155303,PhysRevB.64.245314,PhysRevB.69.241301,Spehner_1998,PhysRevE.67.065202,inverse_B_billiards:arxiv:1911.089144,PhysRevB.71.075331,PhysRevB.77.081403}.
For the interested reader, we suggest Nogaret's excellent review work on the electron dynamics in inhomogeneous magnetic fields~\cite{Nogaret_2010}. 

To understand the quantum behavior of the electron motion, the semiclassical approximation is proved to be a powerful tool.  
It provides a good approximation for calculating the discrete high energy levels and gives a deeper insight into the electron dynamics from the classical point of view. 
From the vast literature, here we refer only to a few classical works~\cite{Landau1981Quantum,Brack:book,heller2018semiclassical}. 
To determine the eigenvalues and eigenstates of a quantum mechanical system, there are two fruitful approaches, namely the originally developed Bohr--Sommerfeld (BS) quantization 
method and the Einstein--Brillouin--Keller (EBK) method, which is an improvement of the BS method. 
The theoretical works mentioned above demonstrate the power of the semiclassical approach. 
In general, the energy levels obtained from the semiclassical quantization rules agree very well with the numerically exact levels even for the lowest eigenstates. 

\begin{figure}[!htb]
	\centering
	\includegraphics[scale=0.4]{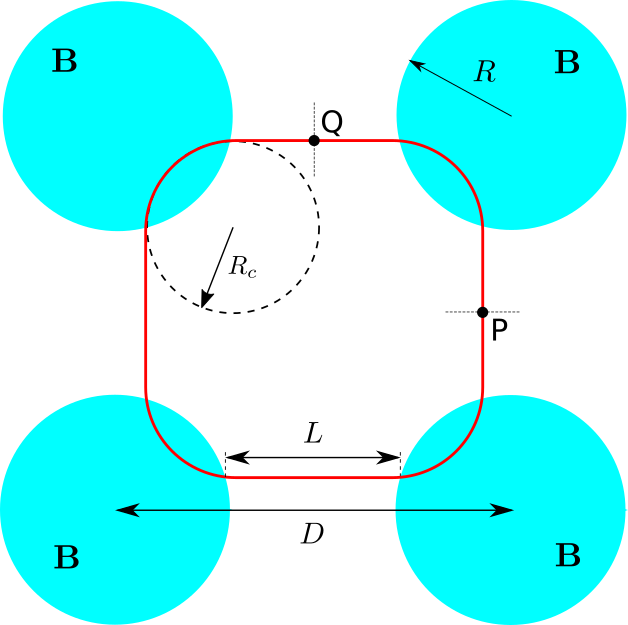}
	\caption{A schematic picture of the system with four magnetic disks.
		The center of the circles of radius $R$ is at the corners of a square of side length $D$. The cyclotron radius is $R_c$. The length of the straight segments of the periodic orbit (red curve) is $L$.
		Inside the four circles, the magnetic field is constant $\boldsymbol{B}$ 
		and perpendicular to the plane, and it is zero outside of the circles. 
		The Poincar\'e map is calculated between points $P$ and $Q$ located at the midpoint of the straight segments. 
	}
	\label{geo_4disk:fig}
\end{figure}

In this work, we study the classical motion of an electron in a 2DEG system which contains circular
regions with a constant magnetic field inside and perpendicular to the plane, while outside the circular regions the magnetic field is zero.
These magnetic disks are placed in a square array as 
shown in~\fref{geo_4disk:fig} for the four magnetic disks.
In what follows, the magnetic field $B$ is conveniently measured by the dimensionless magnetic flux $\Phi/\Phi_0$, where $\Phi = B R^2 \pi$ is the magnetic flux through one magnetic disk and $\Phi_0 = h/e$ is the flux quantum and  $e>0$ is the magnitude of the electron charge.   
In a typical geometry of 2DEG systems taking a magnetic disk of radius $100~\mathrm{nm}$, the magnetic field is an experimentally reasonable value, namely 
$B= \Phi/(R^2\pi)=  0.13~\mathrm{T} $ for the dimensionless magnetic flux  $\Phi/\Phi_0=1$. 

We assume that the system is in the ballistic regime, i.e., the mean free path of the electron is much larger than the size of the system, and thus the electron scattering by impurities is negligible.  
The electron classically follows straight line segments between the disks
and circular arc inside each disk. 
From the equation of motion, we easily find that in the presence of homogeneous magnetic field the radius $R_c$ of such arcs, i.e., the cyclotron radius is given by 
$R_c = \sqrt{2 m E}/(eB)$,
where $m$ and $E$ are the effective mass and the energy of the electron, respectively. 
In 2DEG, e.g., for GaAs/AlGaAs semiconductor heterostucture, 
the electron mass is $m=0.063\, m_e$, where $m_e$ is the free electron rest mass, and the energy $E$ is the order of 
14~meV~\cite{Carlo-Houten}. 
\Fref{geo_4disk:fig} shows a typical periodic orbit. 
The experimental feasibility of such an arrangement is discussed in Refs.~\cite{Carlo-Houten,Carlo-konyv,Nogaret_2010}. 

In this work, we present results obtained from both semiclassical  methods and compare them to that calculated from the exact numerical quantum calculation.   
To complete our semiclassical study, we finally present
results for the probability current density calculated from
quantum calculations. We shall argue that the current flow
patterns can be understood qualitatively from the corresponding classical trajectories.

\section{Stability analysis of periodic orbits }
\label{stability:sec}

To perform the semiclassical quantization, we need to explore the possible periodic orbits and their stability in this system. 
To this end, in this section, we consider the orbit
shown in~\fref{geo_4disk:fig}.
In order to determine its stability, we introduced a local
phase space coordinate system at each point of this orbit.
We introduce two coordinates describing the infinitesimally small perturbation as follows. 
Denote by $\xi$  and $v_\xi$ the deviation from and the velocity component perpendicular to the closed orbit. 
Determining their values on a straight line segment orthogonal to the
periodic orbit at the point $P$ shown in~\fref{geo_4disk:fig} corresponds to taking a Poincar\'e section in the phase space.
The evolution of the deviation $(\xi,v_\xi)$
during one period (i.e. the Poincar\'e mapping) corresponds to four repetitions of the evolution along the segment between point $P$ and $Q$.
The relation between the deviations $(\xi,v_\xi)$ measured 
at point $P$ and $(\xi',v_\xi')$ measured at $Q$ 
can be written in linear approximation as
\begin{equation}
\left( 
\begin{array}{c} 
\xi'/R \\[1ex] v_\xi'/v \end{array}
\right)
= \mathbf{M}
\left(
\begin{array}{c} \xi/R \\[1ex] v_\xi/v \end{array}
\right),
\end{equation}
where the velocity component $v_\xi$ is normalized by the electron's velocity $v= \sqrt{2E/m}$. 
The condition of stability of the studied orbit is given by 
$\left| \mathrm{Tr}\, \mathbf{(M)} \right| < 2 $~\cite{Brack:book}.
The actual form of $\mathbf{M}$ can be obtained 
by splitting the segment $PQ$ into five parts: 
two straight line segments (one starts at $P$ 
and the other one ends at $Q$), the cyclotron orbit, and the infinitesimal neighborhoods of the two points where the orbit enters and leaves the magnetic disk.
Consequently, $\mathbf{M}$ is obtained as the product of the five  stability matrices corresponding to these segments of the orbit and  
we find
\numparts
\begin{equation}
\setlength\arraycolsep{2ex}
\mathbf{M}=
\left(
\begin{array}{cc}
\mu + \frac{(\mu^2-1)l}{2r_c}  &  r_c+\mu l + \frac{(\mu^2-1)l^2}{4r_c} \\[2ex]
\frac{\mu^2-1}{r_c}            &  \mu + \frac{(\mu^2-1)l}{2r_c} 
\end{array}
\right),
\label{stab_4disk:eq}
\end{equation}
where 
\begin{equation}
\mu =  \frac{r_c-\sqrt{2-r_c^2}}{r_c+\sqrt{2-r_c^2}}, \hspace{3mm} \mathrm{and} 
\hspace{3mm} l= \frac{L}{R} =  d-r_c-\sqrt{2-r_c^2}, \\
\label{m_meredekseg_4kor:eq}
\end{equation}
\endnumparts
$d= D/R$, $r_c = R_c/R$,  and $L$ is the length of the straight segments of the periodic orbit as shown in~\fref{geo_4disk:fig}.
The closed orbits exist when $r_c = R_c/R \le \sqrt{2}$.

Now using the stability matrix $\mathbf{M}$ given by~\eref{stab_4disk:eq} 
the stability condition 
$\left| \mathrm{Tr}\, (\mathbf{M}) \right| \le2 $
can be written as 
\begin{equation}
\frac{d}{2} <
r_c+\frac{1}{\sqrt{2-r_c^2}}, \hspace{3mm} \mathrm{and} 
\hspace{3mm} r_c < \sqrt{2}.
\label{stab_cond_4disk:eq}
\end{equation}

\begin{figure}[!htb]
	\centering
	\includegraphics[scale=0.4]{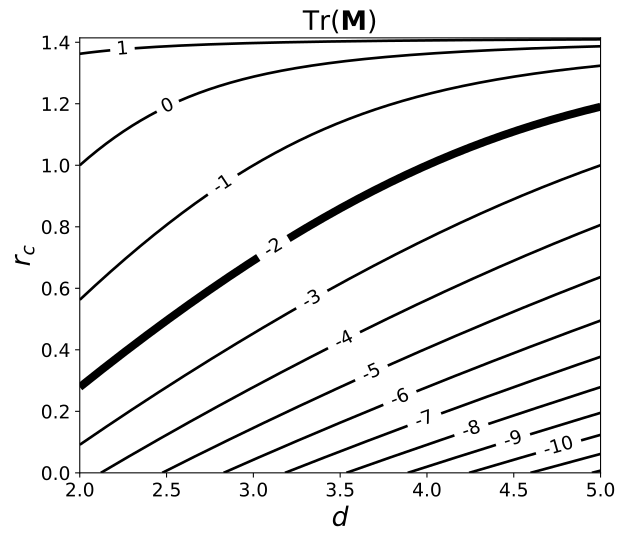}\hspace{5mm}
	\includegraphics[scale=0.4]{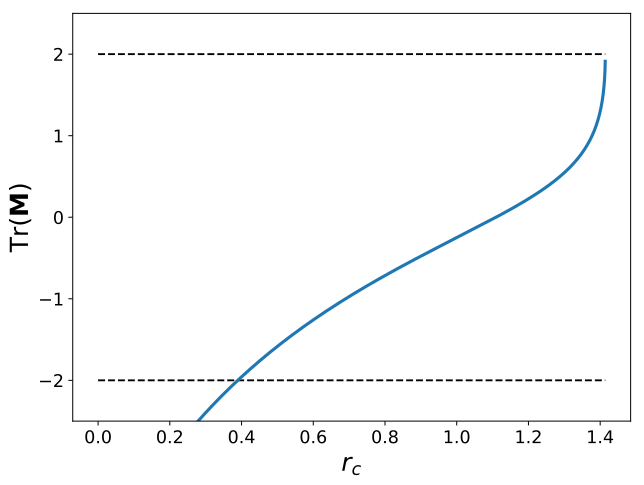}
	\caption{Stability map. The contour plot of  $\mathrm{Tr}\,\mathbf{(M)}$ as functions of the cyclotron radius $r_c = R_c/R$ and $d=D/R$ (left panel) and as a function of $r_c$ for $d=2.25$ (right panel). 		
		On the right panel the stability condition in $\mathrm{Tr}\,\mathbf{(M)}$ is indicated  by dashed lines.}
	\label{stab_con-4disk:fig}
\end{figure}

To get deeper insight into the stability of the periodic orbits we calculated  
$\mathrm{Tr}\, \mathbf{(M)}$ as a function of the cyclotron radius $r_c = R_c/R$ and the separation of the magnetic disks $d=D/R$. 
The numerical results are shown in~\fref{stab_con-4disk:fig}. 
As can be seen from this figure (left panel),  the thick black contour line corresponding to $\mathrm{Tr}\, \mathbf{(M)}=-2$ determines the lower limit of $r_c$ as a function of $d$ for which the periodic orbit is stable. 
This lower limit is implicitly given by the first
inequality in~\eref{stab_cond_4disk:eq} 
taking it as equality, while the upper limit is $r_c = \sqrt{2}$. 
It is clear that the lower limit of $r_c$ increases as the separation $d$ increases, and therefore, the stability region becomes narrower. 
We take $d=D/R=2.25$, for which the stability region is reasonably wide as the magnetic field (scaled by $r_c$) is varied.  
Solving numerically~\eref{stab_cond_4disk:eq} for $d=2.25$, 
we find that the stable region is 
$ 0.38945 \lessapprox r_c < \sqrt{2}$, as can be seen in the right panel 
in~\fref{stab_con-4disk:fig}.  

To see how the periodic orbits become stable, we calculated the two-dimensional Poincar\'e sections at each period.
\begin{figure}[!htb]
	\centering
	\includegraphics[scale=0.5]{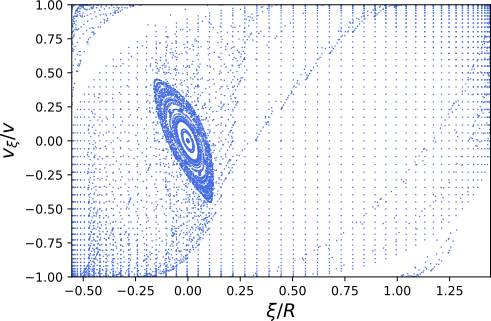}\hspace{5mm}
	\includegraphics[scale=0.5]{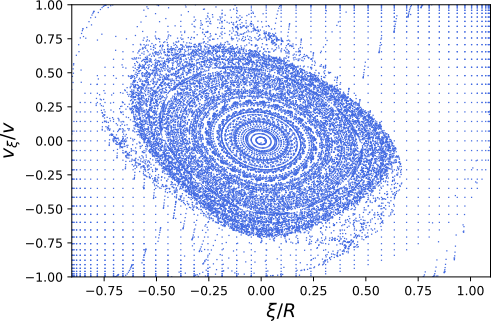}
	\caption{The Poincar\'e surface of section for $r_c=0.45$ 
		(left  panel) and $r_c=0.9$ (right panel), with $d=2.25$.}
	\label{Poinc:fig}
\end{figure}  
Our results are plotted in~\fref{Poinc:fig}
for two different magnetic field strengths.
The left panel shows the emergence of the stable periodic orbit and the surrounding stable island at $r_c=0.45$, since
this value of $r_c$ is slightly larger than the lower limit of the
stability range found in~\fref{stab_con-4disk:fig} for $d=2.25$.
Then, increasing the cyclotron radius $r_c$
(decreasing the magnetic field),
the stable island becomes more characteristic in a region of $r_c$,
as it can be seen in the right panel of~\fref{Poinc:fig} for $r_c=0.9$.
Later the stable island becomes smaller again
and disappears at $r_c=\sqrt{2}$.

Note that from a similar calculation we find that those periodic orbits which touch only two magnetic disks are unstable; thus, we ignore them hereinafter.  
We also carried out the stability analysis for the triangular lattice of magnetic disks and find that the stability region of the Poincar\'e surface of section is smaller than that for the square lattice case. Therefore, from now on, we shall focus only on this latter case.   

\section{Semiclassical quantization}
\label{semiclass_1:sec}

\begin{figure}[!htb]
	\centering
	\includegraphics[scale=0.4]{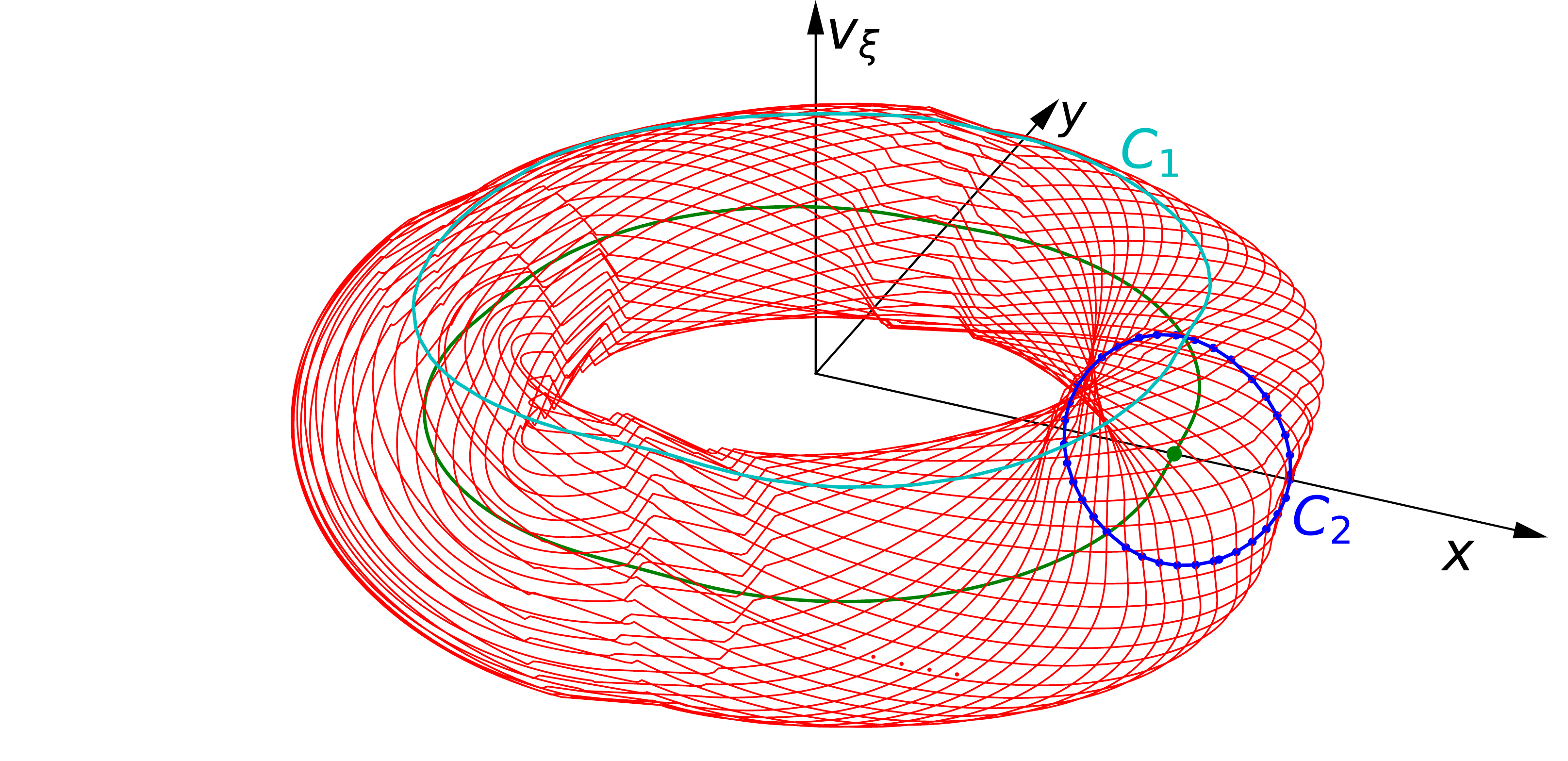}
	\caption{Torus quantization.
		For the representation of the torus, the coordinates $x, y$ of the electron
		and its velocity component $v_\xi$ are used.
		The periodic orbit is drawn in green,
		and $C_1$, $C_2$ are the closed curves on the torus
		along which the integrals are evaluated. 
		\label{EBK_torus:fig}}    
\end{figure}

In case of a stable periodic orbit, a trajectory started in its neighborhood fills up the surface of a two-dimensional torus embedded in the four-dimensional phase space.  
Owing to the energy conservation, such tori can be well represented in a three-dimensional subspace spanned by the real space $(x, y)$ coordinates of the electron and its velocity component $v_\xi$. 
It is well known that the EBK torus quantization can be formulated as~\cite{Brack:book,heller2018semiclassical} 
\numparts
\begin{eqnarray}
\oint_{C_1} \boldsymbol{p}~\mathrm{d}\boldsymbol{r} &=&
\oint_{C_1} 
\left(m \dot{\boldsymbol{r}}-e\boldsymbol{A}\right)~\mathrm{d}\boldsymbol{r} = 
 h\, n_1,
\label{EBK_quant_gen_1:eq} \\[2ex]
\oint_{C_2} \boldsymbol{p}~\mathrm{d}\boldsymbol{r} &=& 
\oint_{C_2} \left(m \dot{\boldsymbol{r}}-e\boldsymbol{A}\right)~\mathrm{d}\boldsymbol{r}= 
 h \left(n_2+\frac12\right),
\label{EBK_quant_gen_2:eq}
\end{eqnarray}
\endnumparts
where the line integrals are along the closed curves $C_1$ and $C_2$, the two topologically independent closed paths, on the torus. 
In our system, the curves are obtained from numerical calculations and shown in~\fref{EBK_torus:fig}. 
Here, $n_1$ and $n_2$ are integer numbers, and 
in magnetic field the canonical momentum of the electron is 
$\boldsymbol{p} = m \dot{\boldsymbol{r}}-e\boldsymbol{A}$, where $\boldsymbol{A}$ is the vector potential related to the magnetic field as  $\boldsymbol{B} = \boldsymbol{\nabla}\times \boldsymbol{A}$. 
The presence of breakpoints in the trajectory forming the torus is a
consequence of the discontinuous change of the magnetic field.
The orbit near the periodic orbit enters the magnetic region earlier or
later than the periodic orbit and starts to change direction with a time
delay compared to it.
Note that 
when the quantization rule (\ref{EBK_quant_gen_2:eq}) is omitted, namely,  when loop $C_2$ is shrunk to the periodic orbit, the original
Bohr--Sommerfeld quantization is recovered, as formulated in~\eref{EBK_quant_gen_1:eq}.

\subsection{Bohr--Sommerfeld quantization of periodic orbits }
\label{Bohr-Sommerfeld:sec}

First, we consider the BS quantization rule given by~\eref{EBK_quant_gen_1:eq}. 
Using Stokes's theorem, the integral here can be written as 
\begin{eqnarray}
\oint_{C_1} \left( m \dot{\boldsymbol{r}}-e\boldsymbol{A}\right)~\mathrm{d}\boldsymbol{r} &= & 
\oint_{C_1}  m \dot{\boldsymbol{r}}~\mathrm{d}\boldsymbol{r} 
-e \, \Phi_{\mathrm{tot}}= h n_1,
\label{BS_Phi_tot:eq}
\end{eqnarray}
where $\Phi_{\mathrm{tot}}$ is the total magnetic flux enclosed by the periodic orbit (along curve $C_1$).
Note that the line integral over a closed path is independent of the gauge of the vector potential. 
The remaining integral can be calculated analytically, 
and finally,~\eref{BS_Phi_tot:eq} becomes: 
\begin{equation}
\frac{\Phi}{\Phi_0}=\frac{n \pi}
{4dr_c + (\pi -2)r_c^2-2r_c\sqrt{2-r_c^2}-4\arcsin{\frac{r_c}{\sqrt{2}}}}, 
\label{BS_quant_4disk:eq}
\end{equation}
where  $\Phi_{\mathrm{tot}}$ is  expressed in terms of the magnetic 
flux $\Phi = B R^2 \pi$  through one magnetic disk, and $n$ is an integer number (for simplicity, here we replace $n_1$ by $n$). 
For a given $r_c$, the energy level $E_n$ can be expressed in terms of the flux $\Phi$ as  
\begin{equation}
\varepsilon_n =  \frac{E_{n}}{E_d} = 4 r_c^2\, {\left(\frac{\Phi}{\Phi_0}\right)}^2, 
\label{epsi_n_BS:eq}
\end{equation}
where $E_d = \frac{\hbar^2}{2m R^2}$ is a characteristic energy scale. 
Now, the energy $\varepsilon_n $ as a function of the magnetic field can be calculated numerically as follows. 
For a given $r_c$ (corresponding to a given magnetic field $B$) and quantum number $n$ one can determine $\Phi/\Phi_0$ and 
$\varepsilon_n $ from equations~(\ref{BS_quant_4disk:eq}) and (\ref{epsi_n_BS:eq}), respectively, and then $\varepsilon_n $ is plotted as a function of $\Phi/\Phi_0$. 
Later on, in chapter~\ref{EBK:sec}, we present numerical results obtained from the BS quantization and the EBK method. 

Finally, it is worth mentioning that, using equations~(\ref{BS_quant_4disk:eq}) and (\ref{epsi_n_BS:eq}),  
one can derive an asymptotic expression for the energy levels $\varepsilon_n$ 
in the limit of large magnetic field, i.e., for $B \to \infty $ ($R_c \to 0$):  
\begin{equation}
\varepsilon_n \to \left(\frac{n \pi}{2d-2\sqrt{2}}\right)^2 \sim n^2, \hspace{3mm} \mathrm{as} \hspace{3mm} B \to \infty ,
\label{BS_approx_anal:eq} 
\end{equation}
which is clearly independent of the magnetic field.
This result can be understood from the fact that the periodic orbit in the limit 
$B \to \infty $  does not penetrate the magnetic disks, only bounces off.

\subsection{EBK quantization}
\label{EBK:sec}

In order to obtain more accurate approximations for the energy levels
than from the BS quantization, we now consider the EBK quantization rules given by (\ref{EBK_quant_gen_1:eq}) and (\ref{EBK_quant_gen_2:eq}). 
The line integrals in these equations are calculated numerically.
For this purpose, the curve $C_2$ is taken as the Poincar\'e section of the torus, i.e., the cross-section of the torus with the plane crossing the periodic orbit perpendicularly at point $P$ 
shown in~\fref{geo_4disk:fig}.
A trajectory from a chosen initial point is followed until its
intersections with this plane are dense enough, 
and the line integral in~\eref{EBK_quant_gen_2:eq} is approximated using the values of
$\boldsymbol{p}$ at the intersections.
The curve $C_1$ is constructed by taking the cross-section of the torus at
subsequent points of the periodic orbit.
Each such cross-section is constructed again as the set of intersections of a long trajectory. 
A point of $C_1$ in each cross-section is chosen as follows. 
Two intersection points are taken, for which $\xi$ is closest to zero with opposite sign and $v_\xi>0$.
Then a linear interpolation is performed to find the point between them at
$\xi=0$.
The succession of these chosen points is used to represent the curve $C_1$ and to calculate the line integral in~\eref{EBK_quant_gen_1:eq}.
Similarly to BS quantization, we can use~\eref{BS_Phi_tot:eq}. 
The line integral of $\dot{\boldsymbol{r}}$ is taken over $C_1$ instead of the periodic orbit; however,  $\Phi_{\mathrm{tot}}$ is unchanged. 
To satisfy the conditions (\ref{EBK_quant_gen_1:eq}) and~(\ref{EBK_quant_gen_2:eq})
for a given pair of $(n_1,n_2)$ values, 
an initial condition is determined by an iterative method to meet those
conditions at the same time, evaluating the line integrals by the above method in each iterative step.
Finally, the energy value immediately follows from the initial velocity.

\begin{figure}[!htb]
	\centering
	\includegraphics[scale=0.45]{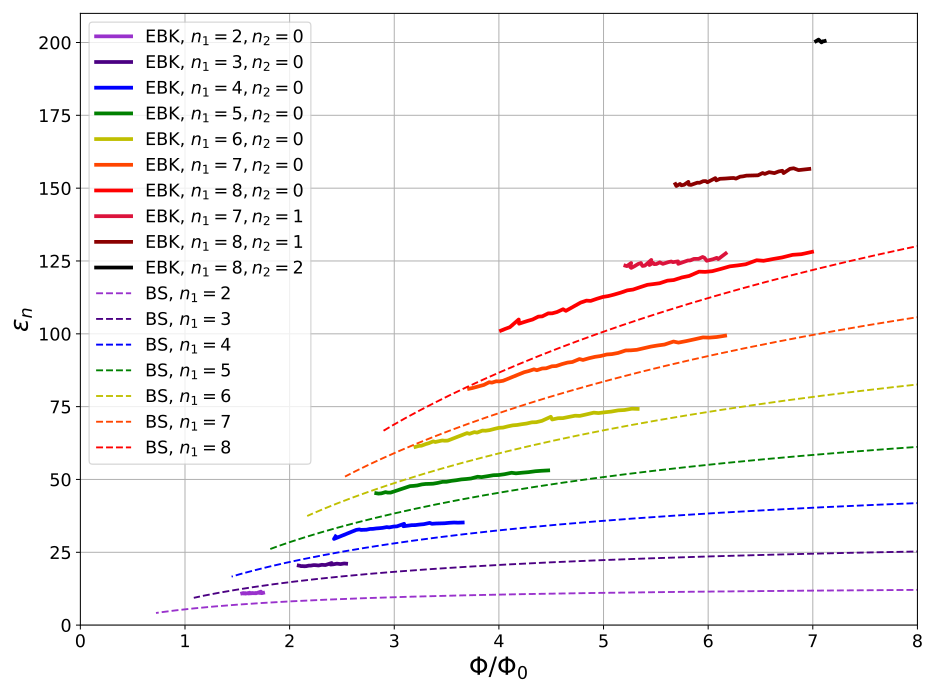} \hspace{5mm}
	\includegraphics[scale=0.5]{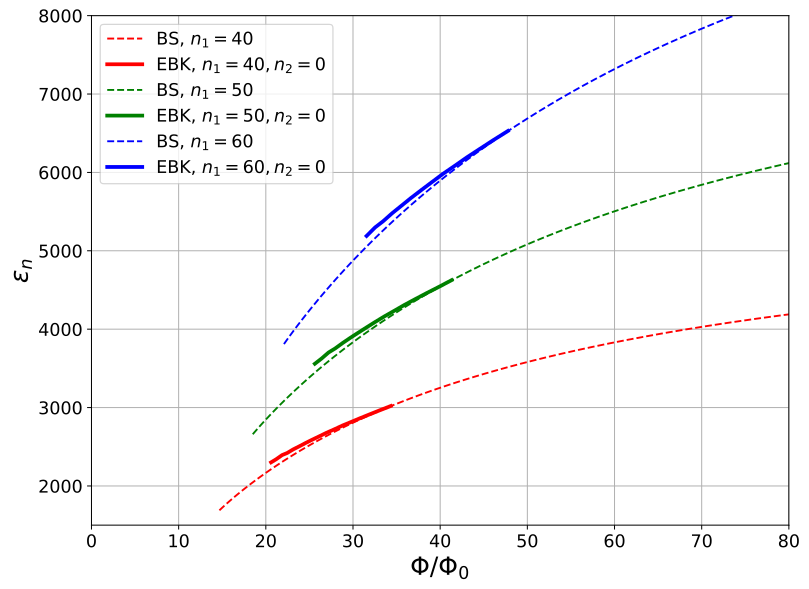}
	\caption{Energy levels $\varepsilon_n$ calculated from the EBK 
		and the BS quantization methods as a function of magnetic fields given by $\Phi/\Phi_0$ 
		for small (upper panel) and large (lower panel) quantum numbers $n_1$ (both for BS and EBK)  and for small $n_2$ (only for EBK), with $d=2.25$.  
		\label{EBK_energy_levels:fig}}    
\end{figure}

As a comparison,~\fref{EBK_energy_levels:fig} shows the energy levels as a function of the magnetic field obtained from the EBK and the BS methods 
(using equations~(\ref{BS_quant_4disk:eq}) and (\ref{epsi_n_BS:eq}))
for small and large quantum number $n_1$, and small number of $n_2$. 

One can see that the numerical results from the two methods agree well 
for all values of the magnetic field as the quantum number $n_1$ increases. 
Note that the allowed values of the magnetic field for the BS method 
are not limited, while this is not the case for the EBK method, since the stability island shrinks (and even shrinks to the loop $C_1$ at some value of the magnetic field), and thus the torus obeying the EBK quantization conditions disappears.
To see how effective the semiclassical quantization is, we compare these results with that obtained from numerically exact quantum calculations. 

\section{Exact quantum calculation}
\label{exact_QM:sec}

In the quantum mechanical treatment, we start from the Hamiltonian of the electron 
given by 
\begin{eqnarray} 
H &=& \frac{{\left(\boldsymbol{p} +e\, \boldsymbol{A}\right)}^2}{2m}.
\label{ham:eq}
\end{eqnarray}
Here  $\boldsymbol{p} = -i \hbar \boldsymbol{\nabla}$ is the canonical momentum operator in position representation, and 
the vector potential $\boldsymbol{A}(\boldsymbol{r})$ for the four magnetic disks in symmetric gauge~\cite{Landau1981Quantum,Sakurai1993Modern:book,Schwabl_QM:book} is given by 
\begin{equation}
\boldsymbol{A}(\boldsymbol{r}) = \frac{1}{2}\, 
\sum_{\substack{i: \boldsymbol{r} \in \mathcal{D}_i}} \boldsymbol{B}\times\left(\boldsymbol{r} - \boldsymbol{c}_i\right) + \frac{R^2}{2}\, 
\sum_{\substack{i: \boldsymbol{r} \notin \mathcal{D}_i}} \frac{\boldsymbol{B}\times\left(\boldsymbol{r} - \boldsymbol{c}_i\right)}{\left|\boldsymbol{r} - \boldsymbol{c}_i\right|^2}\;,
\label{eqvecpot3}
\end{equation}
where the sets $\mathcal{D}_i$ (for $i\in\{1,2,3,4\}$) are the four closed disks, with origins $\boldsymbol{c}_i$, in which the magnetic field is non-zero. 
Since in our 2DEG system the Zeeman energy, which removes the spin degeneracy, can usually be ignored below $B \approx 1$~T, we assume the twofold spin degeneracy~\cite{Carlo-Houten}.  

The energy levels $E_n$ of the bound states are the solutions of the eigenvalue problem $H \psi_n = E_n \psi_n$, where the eigenfunctions  $\psi_n(\boldsymbol{r})$ should tend to zero for $|\boldsymbol{r}| \to \infty $ due to the normalization condition. 
This eigenvalue problem cannot be solved analytically. 
Therefore, we used numerical methods to find the energy eigenvalues and the corresponding wave functions. 
To this end, we discretized the system using a two-dimensional lattice and solved the eigenvalue problem by numerical diagonalization. 
The usual method is to replace the derivative of the wave function by its finite difference~\cite{Burden1989,dahlquist2008numerical_I}. 
Moreover, the vector potential is taken into account 
by the usual Peierls substitution~\cite{Peierls1933,PhysRevB.14.2239,Solyom_book_II:konyv}.  
A nice review of this discretization procedure is given by 
Wimmer in his thesis~\cite{Wimmer:PhD}.

In our numerical work, the wave function is calculated on a square lattice with $(2 M+1)^2$ grid points, where  $M=150$.  
We assume that at the boundary of the calculated region, i.e., at the edges of the square, the wave function is sufficiently small, approximately taken to be zero. 
Thus, we are interested in those energy levels for which the corresponding wave functions are confined to the neighborhood of the relevant periodic orbit.
This confinement was quantified by calculating the sum of the squared modulus of the wave function at the lattice points nearest to the edges of the square. 
Throughout the calculation, the radius of the magnetic disks was $R/a = 60 = 0.4 M$, where $a$ is the lattice constant of the square grid. 
The separation between the neighboring magnetic circles was $D/a  = 135$; thus, $d=D/R=2.25$.  
\begin{figure}[!htb]
	\centering
	\includegraphics[scale=0.4]{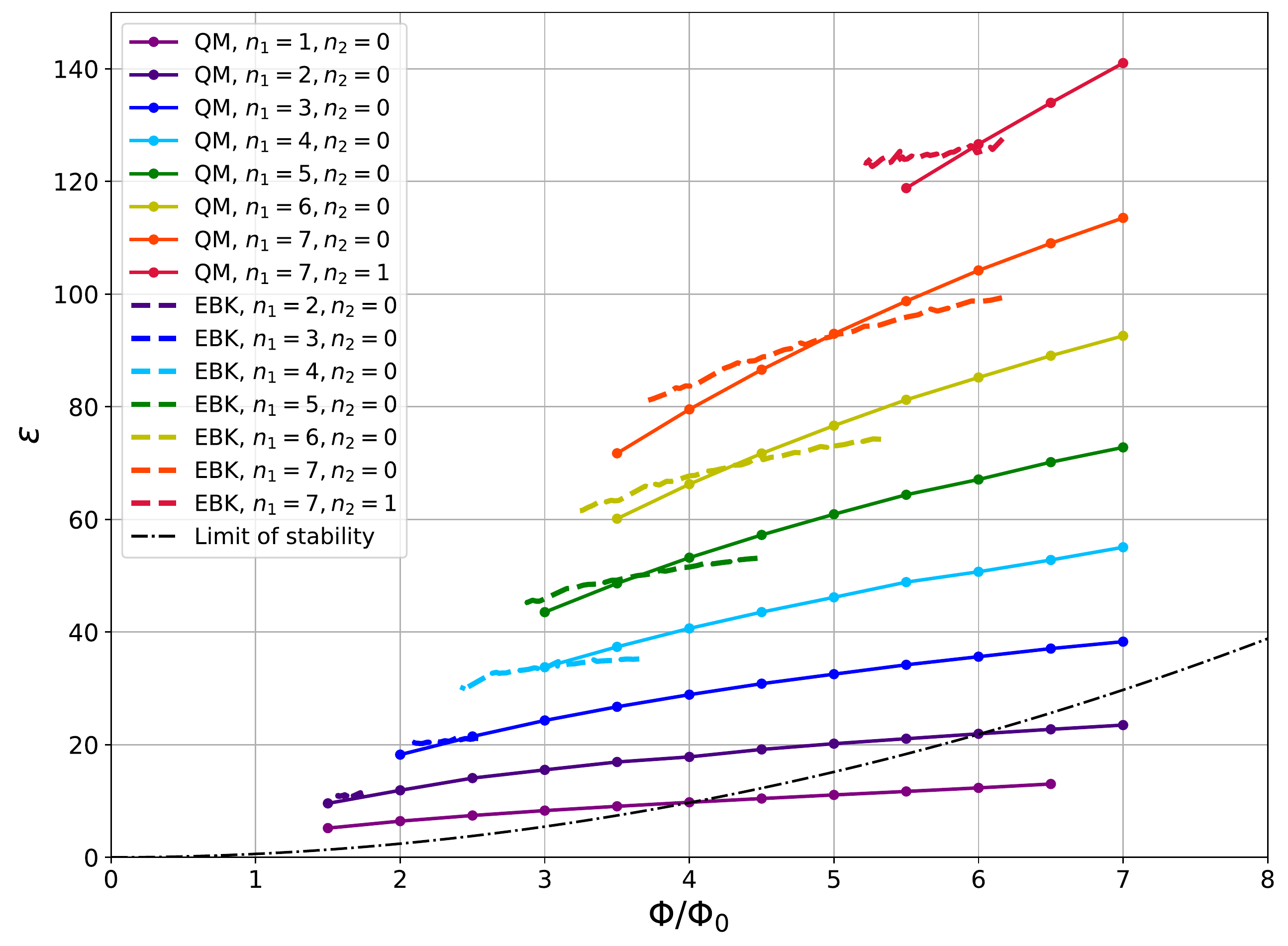}
	\caption{Energy levels $\varepsilon_n$  obtained from EBK quantization (EBK) and exact quantum calculations (QM) for $d=2.25$ as a function of magnetic fields given by $\Phi/\Phi_0$. 
	The energies above the dashed-dotted line (limit of stability) are related to the classically stable periodic orbits.
	In QM calculations the parameters are $M =150$,  $R  =  60$ and $D  =  135$ (so $d=D/R=2.25$), and the value of $\Phi/\Phi_0$ is sampled by 1/2 (the lines are only a guide to the eyes). 
	For details, see the text.   
		\label{energy_levels_QM-EBK_Phi:fig}}    
\end{figure}
Figure~\ref{energy_levels_QM-EBK_Phi:fig} shows the magnetic field ($\Phi/\Phi_0$) dependence of the energy levels obtained from the EBK quantization and from the numerical solutions of the Schr\"odinger equation. 
Note that in the quantum calculations (QM) the quantum numbers 
$n_1$ and $n_2$ are deduced from the current distribution and the phase of the wave functions, as we explain in detail below.  
In addition, we plotted the curve of the limit of stability (dashed-dotted line) obtained from~\eref{epsi_n_BS:eq},  
with $r_c = 0.38945$ as the lower limit of the stability condition calculated  from the first inequality in~\eref{stab_cond_4disk:eq} for $d=2.25$.     
One can see that the EBK quantization is a reasonable good approximation 
of the exact quantum calculations, even for quite low energy levels. 
Although, the magnetic field dependence of the energy levels is slightly  different in the two results, namely the slope of the energy levels as a function of $\Phi/\Phi_0$ is smaller in the semiclassical calculations  compared to the quantum mechanical results. 

We should emphasize two more features in our results. 
The first one is related to the fact that for all calculated states with
$n_2=0$, there is no solution from the EBK quantization rule below some
critical values of the magnetic field.
We found that these lower critical values of the magnetic field agree well
with the values for which the solution of the exact quantum calculation
disappears.
The second feature can be observed when we increase the magnetic field
until there is no solution from the EBK method for given quantum numbers 
$n_1$ and $n_2$.
Beyond this upper critical value of the magnetic field, the EBK method does not work because the stability island is too small to form the necessary torus, though the periodic orbit is still stable. 
However, the exact quantum calculation in such cases
still gives quantum states corresponding to the selected quantum numbers $n_1$ and $n_2$.
Furthermore, the quantum state exists in an interval of values of the magnetic field where the periodic orbit is unstable (see, e.g., $4 \lessapprox \Phi/\Phi_0 \lessapprox 6.5$ for $n_1=1$ and $n_2=0$, and 
$6 \lessapprox \Phi/\Phi_0 \le 7$ for $n_1=2$ and $n_2=0$  
in~\fref{energy_levels_QM-EBK_Phi:fig}). 
The reason why there is a different behavior of the energy levels when decreasing or increasing the magnetic field can be related to the different kind of change in the classical dynamics. 
Namely, when decreasing the magnetic field, the stability island shrinks to the periodic orbit and it disappears together with the
stability island.
In contrast, when increasing the magnetic field, the stability island shrinks to the periodic orbit and disappears, but the periodic orbit itself remains, only looses its stability.

\section{Current distribution}
\label{current_dist:sec}

So far, starting from the classical dynamics of an electron in the system, we calculated the approximated energy levels using the BS and EBK quantization rules. 
Now, in this section, we follow the opposite approach. 
Considering the exact quantum state and the current density distribution, we establish the underlying classical dynamics.   
As we shall see, the emergence of the classical dynamics can be observed directly by plotting the wave functions and the corresponding current density distributions.  
In this way, we have additional insight into the electron dynamics 
in a classical picture.
To this end, not only the eigenvalues but also the corresponding eigenfunctions of the Hamiltonian are calculated for a few energy levels. 
Then, from the eigenstates $ \psi$, we obtain the probability current density. 

The probability current density~\cite{Landau1981Quantum,RevModPhys.57.339,PhysRevB.43.4179,Schwabl_QM:book,Sakurai1993Modern:book} for an electron in magnetic field is given by 
\begin{eqnarray}
\boldsymbol{J}(\boldsymbol{r}) &=& -\frac{i\hbar}{2m}
\left[
\psi^*\boldsymbol{\nabla}\psi - \psi \boldsymbol{\nabla}\psi^* \right]
+\frac{e}{m}\boldsymbol{A}(\boldsymbol{r})|\psi(\boldsymbol{r})|^2 . 
\label{current_1_def:eq}
\end{eqnarray}
Note that writing the wave function in the form $\psi(\boldsymbol{r}) = \left| \psi(\boldsymbol{r})\right| e^{i \chi(\boldsymbol{r})}$, 
the current can be expressed alternatively as 
\begin{equation}
\boldsymbol{J}(\boldsymbol{r})  = 
\frac{\hbar\, \boldsymbol{\nabla}\chi(\boldsymbol{r}) +e \boldsymbol{A}(\boldsymbol{r})}{m} {\left|\psi(\boldsymbol{r})\right|}^2,  
\label{current_2:eq}
\end{equation}
valid everywhere except at points where the phase $\chi(\boldsymbol{r})$ 
of the wave function is discontinuous, i.e., its value jumps 
$\pi$ or $2\pi$. 
To extend the validity of equation \eref{current_2:eq} in agreement with 
\eref{current_1_def:eq}, one can define a vector field  $\boldsymbol{P}(\boldsymbol{r})$ as 
\begin{equation}
\boldsymbol{P}(\boldsymbol{r}) =  \mathrm{Im}\left\{\frac{\boldsymbol{\nabla}
	\psi(\boldsymbol{r})}{\psi(\boldsymbol{r})}\right\}. 
\label{current_P_def:eq}
\end{equation}
Mathematically, this is a continuous extension of 
$\boldsymbol{\nabla} \chi(\boldsymbol{r})$.  
It defines the quantum mechanical \textit{streamline} which is everywhere parallel to the local vector $\boldsymbol{P}(\boldsymbol{r})$ 
(see a series of papers on this topic by Hirschfelder~\cite{Hirschfelder_I:cikk,Hirschfelder_II:cikk,Hirschfelder_III:cikk,Hirschfelder_IV:cikk,Hirschfelder_1977:cikk,SOSKIN2001219} and by Berry~\cite{Berry_438_2011,berry501_2017}). 

Note that $\boldsymbol{P}(\boldsymbol{r})$  in~\eref{current_P_def:eq} is not defined at the nodes where $\left|\psi(\boldsymbol{r})\right| = 0$.
However, it is easy to see that  
$\boldsymbol{\nabla}\times \boldsymbol{P}(\boldsymbol{r}) = 0 $, 
if $\left|\psi(\boldsymbol{r})\right| \neq 0$, and then the circulation, e.g., the line integral of $\boldsymbol{P}(\boldsymbol{r})$ around any closed loop $C$ not crossing the nodes of the wave function is  
\begin{equation}
\oint_C \boldsymbol{P}(\boldsymbol{r})\, 
\mathrm{d} \boldsymbol{r} = 2\pi N,
\label{P_quant:eq}
\end{equation}
where $N$ is an integer number. 
The circulation is quantized and the streamlines form a quantized vortex  around the nodes of the wave function. 
This equation is an analogy of the semiclassical quantization condition and the direct relation will be discussed below. 

Finally, the current density can be rewritten in terms of the vector field $\boldsymbol{P}(\boldsymbol{r})$ in the following way 
\begin{eqnarray}
\boldsymbol{J}(\boldsymbol{r}) &=& 
\frac{\hbar \, \boldsymbol{P}(\boldsymbol{r}) 
	+e\, \boldsymbol{A}(\boldsymbol{r})}{m}
{\left|\psi(\boldsymbol{r})\right|}^2 . 
\label{current_jP_def:eq}
\end{eqnarray}

To see the manifestation of the classical dynamics in the quantum mechanical results, we selected a few states from those shown in~\fref{energy_levels_QM-EBK_Phi:fig} for which the relation between the EBK quantum numbers $n_1$ and $n_2$ and the phase $\chi(\boldsymbol{r})$ of the exact wave function is obvious. 
Thus, we could demonstrate how the classical mechanics gets reflected in the quantum mechanics. 
In~\tref{abrak:table}, for different magnetic field strengths, our selected states are listed together with the energy levels obtained from the quantum mechanical calculations and from the BS and EBK methods, using the given quantum numbers $n_1$ and $n_2$. 
\begin{table}[h]
	\centering
	\begin{tabular}{||c|c|c|c|c|c||}  \hline
		$(n_1,n_2)$ &$\frac{\Phi}{\Phi_0}$ &  BS & EBK & QM  & Figure \\[2ex]   \hline
		$(3, 0)$ & 2.5 & 16.7 & 20.9 & 21.6 & \ref{Hullamfv8:fig} \\[1ex] \hline
		$(4, 0)$ & 3.5 & 30.5 & 35.0 & 37.4 & \ref{Hullamfv9:fig} \\[1ex] \hline
		$(5, 0)$ & 3.5 & 42.2 & 49.2 & 48.6 & \ref{Hullamfv10:fig} \\[1ex] \hline
		$(6, 0)$ & 4 & 58.9 & 67.7 & 66.3 & \ref{Hullamfv2:fig} \\[1ex]
		\hline
		$(7, 0)$ & 4 & 72.9 & 83.7 & 79.4 & \ref{Hullamfv3:fig} \\[1ex]
		\hline
		$(7, 1)$ & 5.5 & - & 124.0 & 118.8 & \ref{Hullamfv11:fig} \\[1ex] \hline
		$(8, 2)$ & 7.1 & - & 200.4 & 189.4 & \ref{Hullamfv12:fig} \\[1ex] \hline
		$(1,0)$ & 6 & 3.2 & - & 12.3 & \ref{Hullamfv15:fig} \\[1ex]
		\hline
		$(-1,0)$ & 2 & - & - & 3.0 & \ref{Hullamfv13:fig} \\[1ex]
		\hline
	\end{tabular}
	\caption{\label{abrak:table} Energy levels $\varepsilon$ obtained from the BS and EBK quantizations with given quantum numbers $n_1$ and $n_2$, and from the exact quantum calculations (QM) for a given magnetic field strength. 
		In the last column, the figure references for the corresponding quantum states are given.}
\end{table}
The last two states in this table are special, and will be discussed later.

One can see from~\tref{abrak:table} that the BS and EBK method provides a reasonable good approximation of the exact quantum calculation (QM) of the energy levels. 
Furthermore, when both methods can be applied the EBK method gives better
results.
Note that the BS method does not take into account the change of the
wavefunction transversely to the periodic orbit. 
Thereby, it is expected to best approximate the states with $n_2=0$. Correspondingly, the BS results are only entered into the rows of the table with $n_2=0$.

\begin{figure}[!htb]
	\centering
	\includegraphics[width=15cm]{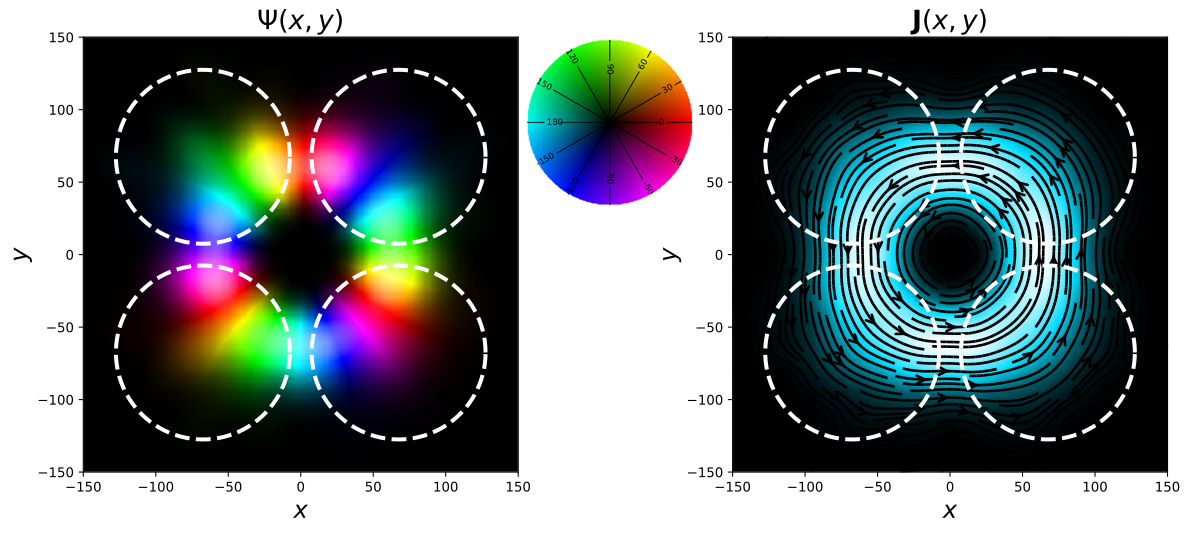}
	\caption{Phase $\chi(\boldsymbol{r})$ of the wave function 
	(left panel) and the corresponding current distribution $\boldsymbol{J}(\boldsymbol{r})$ (right panel) for magnetic field corresponding to $\Phi/\Phi_0 = 2.5$ and energy level $\varepsilon = 21.6$  obtained from the numerical quantum calculation. In the BS and EBK calculations, the quantum numbers are  $(n_1,n_2) = (3,0)$. 
		The shading corresponds to the squared modulus of the wave function.
		The parameters are the same as in~\fref{energy_levels_QM-EBK_Phi:fig}.
		The central panel shows the color circle, i.e., the hue of the color related to the phase 
		$\chi(\boldsymbol{r})$ of the wave function (see the text). 
		\label{Hullamfv8:fig}}    
\end{figure}

\begin{figure}[!htb]
	\centering
	\includegraphics[width=16cm]{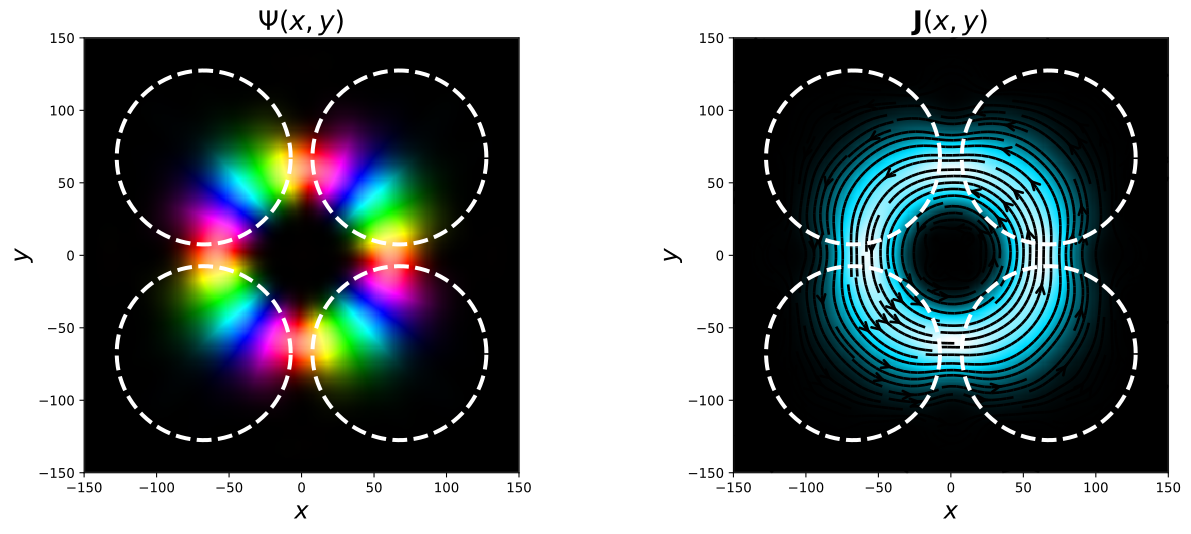}
	\caption{
		Same as in~\fref{Hullamfv8:fig} for $\Phi/\Phi_0 = 3.5$ and 
		energy level $\varepsilon = 37.4$.
		In the BS and EBK calculations the quantum numbers are  $(n_1,n_2) = (4,0)$. 
		\label{Hullamfv9:fig}}    
\end{figure}

\begin{figure}[!htb]
	\centering
	\includegraphics[width=16cm]{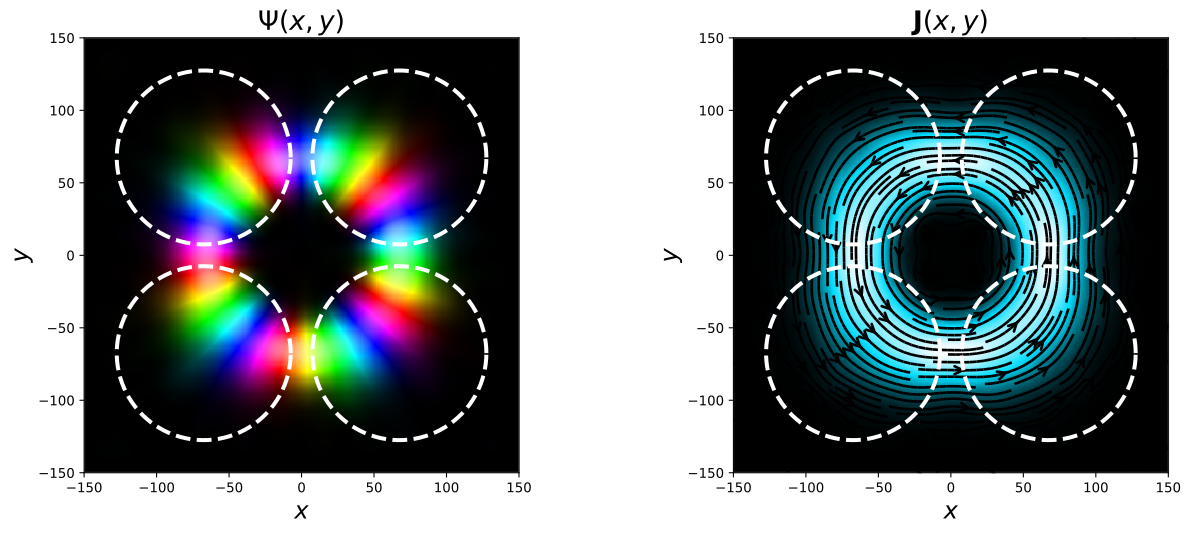}
	\caption{
		Same as in~\fref{Hullamfv8:fig} for $\Phi/\Phi_0 = 3.5$ and 
		energy level $\varepsilon = 48.6$.
		In the BS and EBK calculations the quantum numbers are  $(n_1,n_2) = (5,0)$. 
		\label{Hullamfv10:fig}}    
\end{figure}

\begin{figure}[!htb]
	\centering
	\includegraphics[width=16cm]{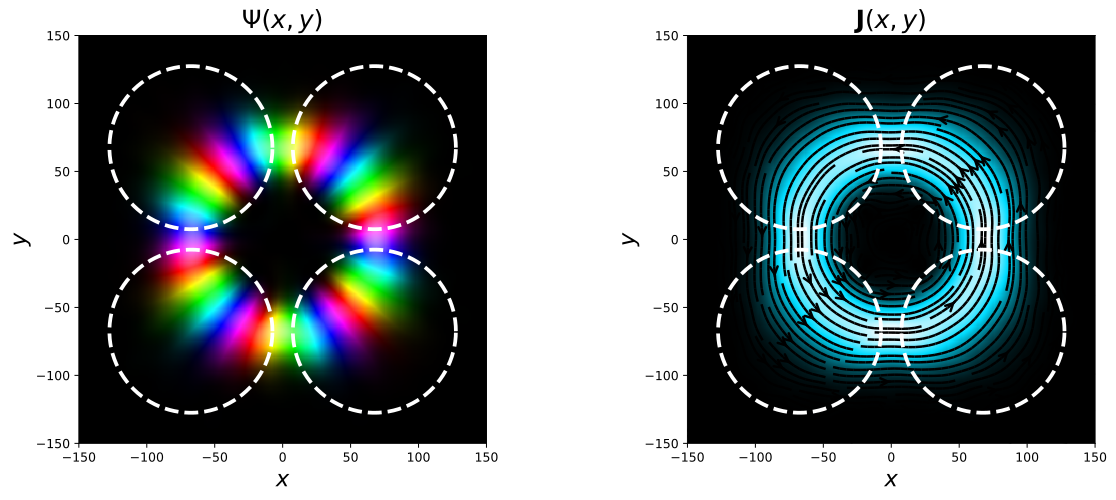}
	\caption{
		Same as in~\fref{Hullamfv8:fig} for $\Phi/\Phi_0 = 4$ and 
		energy level $\varepsilon = 66.3$.
		In the BS and EBK calculations the quantum numbers are  $(n_1,n_2) = (6,0)$. 
		\label{Hullamfv2:fig}}    
\end{figure}

\begin{figure}[!htb]
	\centering
	\includegraphics[width=16cm]{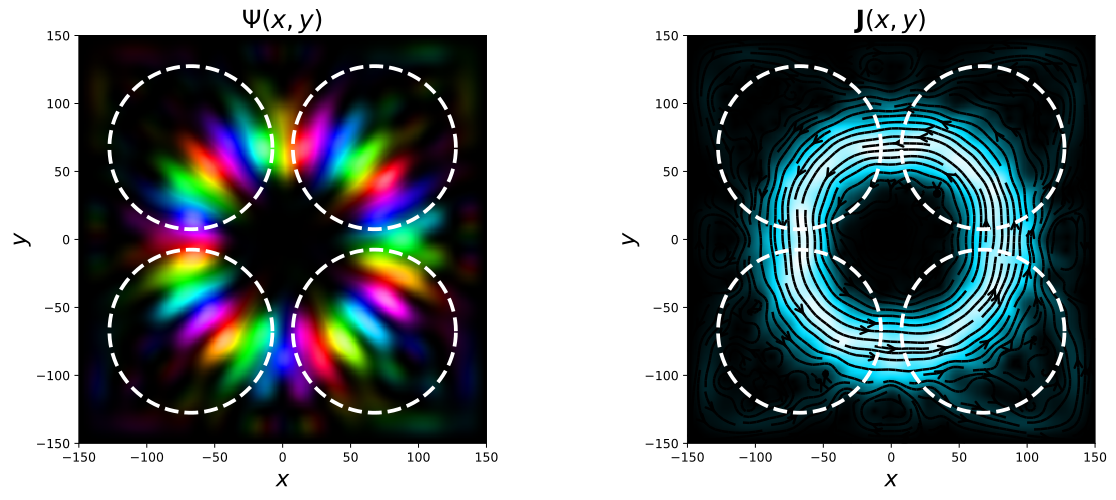}
	\caption{
		Same as in~\fref{Hullamfv8:fig} for $\Phi/\Phi_0 = 4$ and 
		energy level $\varepsilon = 79.4$.
		In the BS and EBK calculations the quantum numbers are  $(n_1,n_2) = (7,0)$. 
		\label{Hullamfv3:fig}}    
\end{figure}

\begin{figure}[!htb]
	\centering
	\includegraphics[width=16cm]{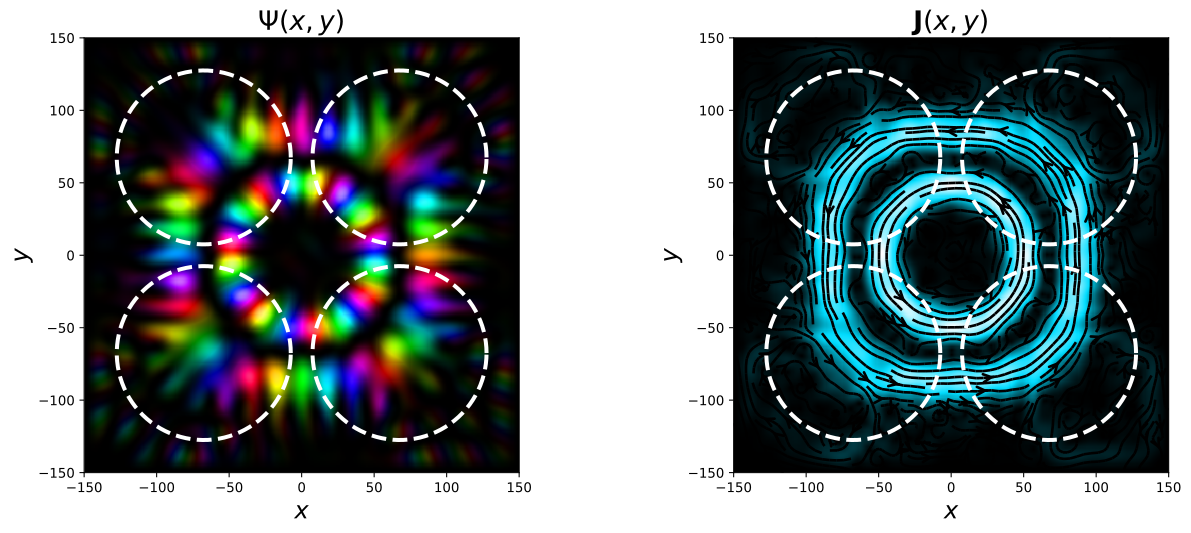}
	\caption{
		Same as in~\fref{Hullamfv8:fig} for $\Phi/\Phi_0 = 5.5$ and 
		energy level $\varepsilon = 118.8$.
		In the EBK calculations the quantum numbers are  $(n_1,n_2) = (7,1)$.
		\label{Hullamfv11:fig}}    
\end{figure}

\begin{figure}[!htb]
	\centering
	\includegraphics[width=16cm]{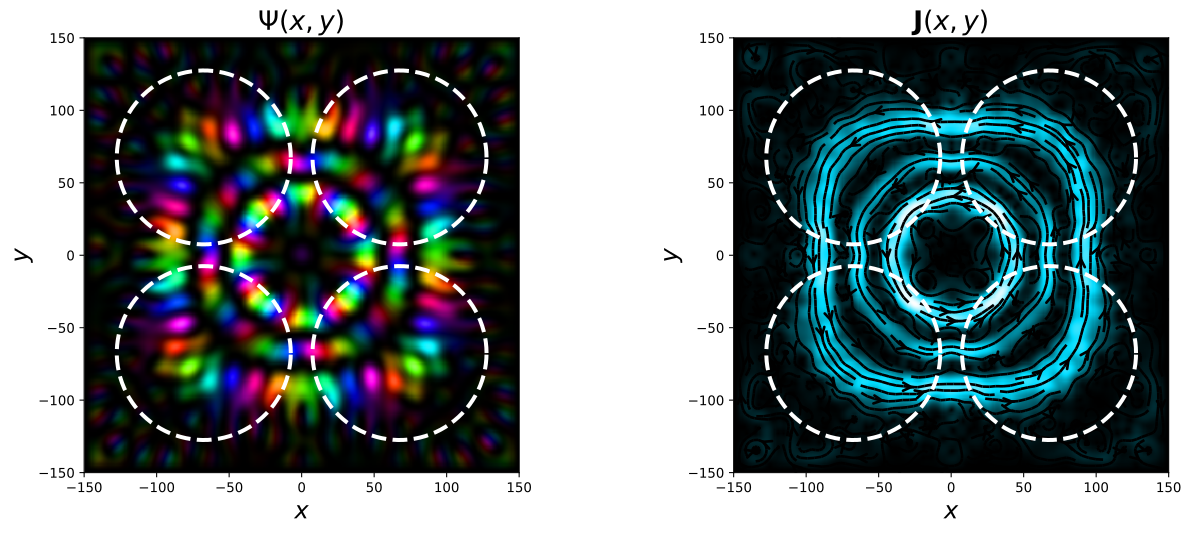}
	\caption{
		Same as in~\fref{Hullamfv8:fig} for $\Phi/\Phi_0 = 7.1$ and 
		energy level $\varepsilon = 189.4$.
		In the EBK calculations the quantum numbers are  $(n_1,n_2) = (8,2)$.
		\label{Hullamfv12:fig}}    
\end{figure}

\begin{figure}[!htb]
	\centering
	\includegraphics[width=16cm]{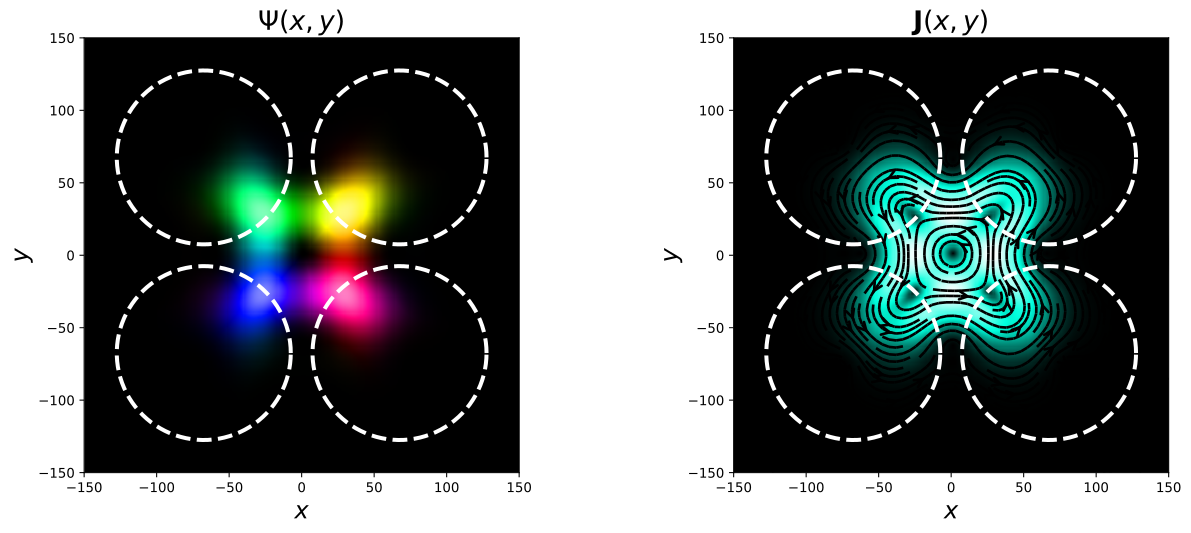}
	\caption{
		Same as in~\fref{Hullamfv8:fig} for $\Phi/\Phi_0 = 6$ and 
		energy level $\varepsilon = 12.3$.
		In the BS calculations the quantum numbers are  $(n_1,n_2) = (1,0)$.
		\label{Hullamfv15:fig}}    
\end{figure}

In figures~\ref{Hullamfv8:fig}--\ref{Hullamfv15:fig} we now show 
the square moduli ${\left|\psi(\boldsymbol{r})\right|}^2$ and the phases  $\chi(\boldsymbol{r})$ of the wave functions and the corresponding current distributions $\boldsymbol{J}(\boldsymbol{r})$ for states listed 
in~\tref{abrak:table}. 
The phase $\chi(\boldsymbol{r})$ is coded by the hue of the color. 
The shading is proportional to the squared modulus of the wave function ${\left|\psi(\boldsymbol{r})\right|}^2$ (for the eyes black corresponds to values of  ${\left|\psi \right|}^2/{\left|\psi \right|}^2_{\mathrm{max}}$ below 0.05).
This colorization method~\cite{Wegert2010PhasePO,wegert2012visual,Thaller_10.5555/556258} 
is used in all subsequent figures. 

One can see from figures~\ref{Hullamfv8:fig}--\ref{Hullamfv15:fig} that the wave functions are well localized around the four magnetic disks with negligible squared modulus of the wave function at the edges of the squared region used in our numerical calculations. 
Moreover, the quantum numbers $n_1$ and $n_2$ associated to the semiclassical quantization rules can be extracted from the phase plot and the current distribution.
Indeed, the quantum number $n_1$ is related to the number how many times
the color goes through the colors of the rainbow (i.e. how many times the
phase $\chi(\boldsymbol{r})$ changes by $2\pi$) when we follow a point  circling ones around the ring.
We should emphasize that this is a manifestation of equation \eref{P_quant:eq} if we perform the integral along the ring; in this case  we can identify $N$ as $N = n_1$. 
Similar behavior of the phase changes can be observed around a specific kind of pole of a complex function~\cite{Wegert2010PhasePO,wegert2012visual,Thaller_10.5555/556258}.
At the same time, the quantum number $n_2$ is related to 
the number of radial nodes of the wave functions, and it is also reflected 
in the current distributions $\boldsymbol{J}(\boldsymbol{r})$.
Namely, the number of rings in the wavefunction plots is the same as the
number of current density loops as a vortex~\cite{Hirschfelder_II:cikk}  around the origin, and they are equal to $ n_2 + 1 $.

Although, in the examples presented above, the clear correspondence between the classical and quantum mechanical results is very spectacular, this is not always the case.    
For instance, the last two quantum states in~\tref{abrak:table}  are rather special ones, as their quantum states cannot be related to any stable classical orbit as those studied before. 

First, consider~\fref{Hullamfv15:fig} (cf.\ last but one row 
in~\tref{abrak:table}). 
To clarify the relation of this state to those we discussed before, 
we should note the following. 
Increasing the value of $\Phi/\Phi_0$, the energy and its wave function change continuously, 
in spite of the fact that the periodic orbit becomes unstable and the stability island disappears at some critical value of $\Phi/\Phi_0$, depending on the quantum numbers $n_1$ 
and $n_2$. 
For instance, in case of the curve labeled by QM, $n_1=1,n_2=0$ 
(lowest curve) in~\fref{energy_levels_QM-EBK_Phi:fig},  
it occurs at $\Phi/\Phi_0  \gtrapprox  4$, where the curve crosses
the limit curve of stability. 
Thus,~\fref{Hullamfv15:fig} shows a quantum state at $\Phi/\Phi_0 = 6$, where the periodic orbit is unstable. 
However, the overall structure of the wave function is similar to what can
be seen for $ \Phi/\Phi_0 < 4$,  only a distortion can be observed
as shown in~\fref{Hullamfv15:fig}.
Such a quantum state localized to an unstable periodic orbit is called scar state in the literature~\cite{heller2018semiclassical}. 
The change of the phase and the lack of radial nodes 
are consistent with the quantum numbers $n_1=1, n_2=0$.
Moreover, the current distribution keeps its ring structure, 
but follows the deformation of the ring shape seen in the probability density.

\begin{figure}[!htb]
	\centering
	\includegraphics[width=16cm]{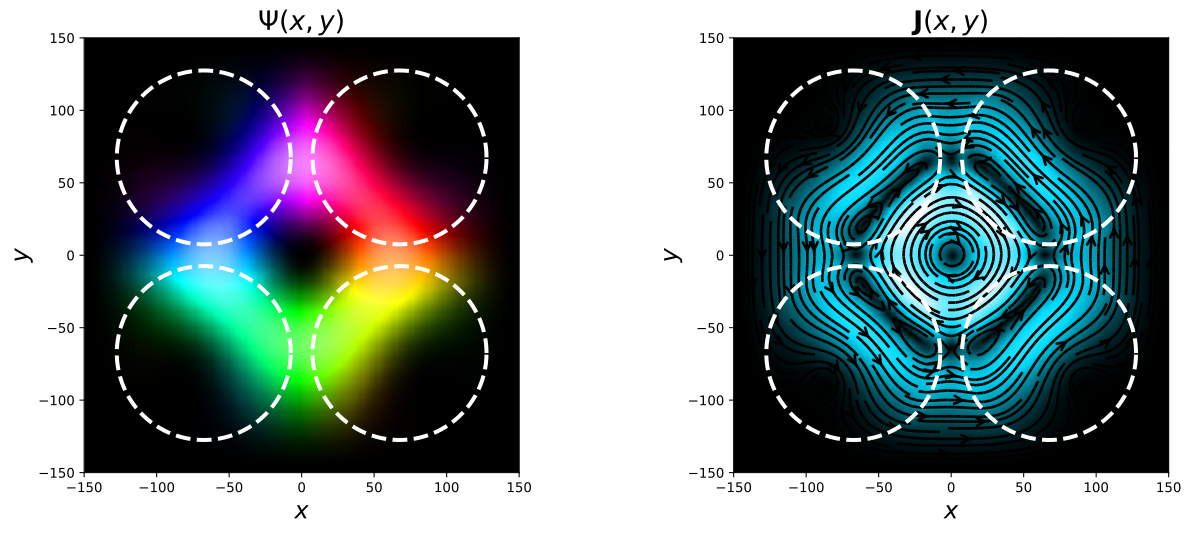}
	\caption{
		Same as in~\fref{Hullamfv8:fig} for $\Phi/\Phi_0 = 2$ and 
		energy level $\varepsilon = 3.0$.
		In the BS and EBK calculations the quantum numbers are  $(n_1,n_2) = (-1,0)$.
		\label{Hullamfv13:fig}}    
\end{figure}

We now consider another special state shown 
in~\fref{Hullamfv13:fig} (cf.\ last row in~\tref{abrak:table})
in which the wave function has such a similar structure that can be understood in the framework of EBK quantization, and to the last mentioned scar state.
Here, it is a striking feature that the order of the colors
in the figure, i.e., the change of the phase of the wave function is opposite to the states shown before. 
According to the rules used to read off the values of the quantum numbers $n_1$ and $n_2$, the opposite order of the colors can be interpreted as a negative value of $n_1$, namely 
$n_1 = -1$, furthermore, since there is only a single ring, 
$n_2 = 0$ in~\fref{Hullamfv13:fig}.  
However, in the current density, two rings can be seen. 
This is a clear consequence of the competition between the contribution of the two terms in~\eref{current_jP_def:eq}. 
The direction of the current density is clockwise in the inner ring
and counterclockwise in the outer ring.

\section{Conclusions}
\label{conlusion:sec}

In this work, we have studied the classical and quantum mechanical dynamics of electrons moving on a two-dimensional plane in an inhomogeneous magnetic field applied perpendicular to the plane. 
We determined the stability conditions of periodic orbits as a function of the magnetic field.
To have a deeper insight into the quantum mechanical behavior of the electron dynamics of our systems, we applied two kinds of semiclassical quantization, namely the Bohr--Sommerfeld and the  Einstein--Brillouin--Keller method. 
To this end, we calculated the two-dimensional Poincar\'e surface of sections for different magnetic field strengths. 
Under given initial conditions we numerically calculated the two-dimensional invariant torus embedded in the four-dimensional phase space.
In the case of the Bohr--Sommerfeld method, we derived an analytical expression for the quantization condition of the energy levels as a function of the magnetic field.
While, applying the Einstein--Brillouin--Keller method, the energy levels were calculated numerically, using the quantization rules on the tori around the stable orbits.

To compare the results obtained from these semiclassical treatments, 
we performed exact quantum calculations, using the discretized version of the corresponding Schr\"odinger equation.  
Moreover, to demonstrate how the classical dynamics emerges 
in the eigenfunctions obtained from the quantum calculations, we determined the current distributions and the phase of the wave function for different magnetic fields. 
Our results obtained from the two semiclassical methods are in good agreement with that found from the quantum calculations. 
From the structure of the current distribution and the phase of the different wave functions, we were able to deduce the two quantum numbers $n_1$ and $n_2$ characterizing the energy levels in the semiclassical methods.  

Finally, we presented two examples when there is no clear correspondence between the classical and quantum mechanical results. 
Indeed, in these cases, quantum states cannot be related directly to any stable classical orbit. 
One example is the so-called scar state, when the quantum state is localized to an unstable periodic orbit. 
In the other example, the order of the colors of the phase 
of the wave function is opposite to that observed in our previous examples. This feature can be interpreted as a negative value of the quantum number $n_1$.
Moreover, the current density is circulating in two
rings with opposite direction. Thus, it is not consistent with the classical motion in the neighborhood of the periodic orbit.

As we have shown, the specially designed inhomogeneous magnetic field 
can be used to confine the electron around the four magnetic dots.  
Then, by attaching contacts to our system, the electric conductance can be measured experimentally. 
Indeed, owing to this confinement, one can expect a drop in the conductance when the electron's energy is close to the energy levels of 
the bound states. 
Thus, varying the magnetic field this electron's localization could be used to control the conductance in  magnetotransport experiments. 
Moreover, our work can also be extended to study the electron's dynamics in other materials, such as graphene~\cite{PhysRevLett.122.137701} or heterostructure with coupled atomically thin layers~\cite{Fulop_2018}.

\ack
We would like to thank L.~Oroszl\'any and J.~Koltai for helpful discussions.
This work was supported by NKFIH within the Quantum Technology National Excellence Program (Project No. 2017-1.2.1-NKP-2017-00001),
by the ELTE Institutional Excellence Program (TKP2020-IKA-05) financed by the Hungarian Ministry of Human Capacities, and Innovation Office (NKFIH) through Grant Nos. K134437.

\appendix

\section{Perturbation theory }
\label{perturb:sec}


In order to separate edge states and magnetically localized states, one can apply perturbation theory. 
Consider a one-parameter ($\delta$) family of rectangular boxes with sides $2L+\delta$ and $2L$ in which the electric potential is given by
\begin{equation}
U_\delta(x,y) =
\left\{ \begin{array}{cl}
\hfill 0, &  ~\mathrm{if}~(x,y)\in[-L,L+\delta]\times[-L,L], \\
\hfill -\infty, & ~\mathrm{if}~(x,y)\notin[-L,L+\delta]\times[-L,L].
\end{array}\right.
\end{equation}
In the presence of magnetic field, the Hamiltonian $H_\delta$ is the same as \ref{ham:eq} inside the box. 
Owing to the potential walls, the wave function must vanish on the boundary of the box. For eigenvalues $E_{\delta,n}$ and eigenvectors $\psi_{\delta,n}$ of $H_\delta$, one can write 
\begin{equation}
\left<\psi_{\delta,n}\Big|\hat H_\delta \psi_{\delta,n}\right> = E_{\delta,n}\;.
\label{eig:eq}
\end{equation}

Our main goal is to study the effect of the side length of the box on the energy levels (or more precisely, the behavior of $E_\delta$ around $\delta=0$). It is reasonable to assume that the magnetically localized states are only weakly affected by $\delta$, whereas the edge states 
are strongly. 
Therefore, this is a useful method for separating the aforementioned states. To this end, we calculate 
$\partial E_{\delta,n}/\partial \delta$. 
From~\eref{eig:eq}, one finds 
\begin{eqnarray}
\label{der:eq}
\frac{\partial E_{\delta,n}}{\partial \delta}
&=&\frac{1}{2m}\partial_\delta\left\{\int_{-L}^{L+\delta}\mathrm{d}x\int_{-L}^{L}\mathrm{d}y~ \psi^*_{\delta,n}\left[(i\hbar\partial_x + q A_x)^2
\right. \right. \nonumber \\[2ex]
&& \left. \left. +(i\hbar\partial_y + qA_y)^2\right]\psi_{\delta,n} \right\}.
\end{eqnarray}
Using Leibniz's integral rule, integration by parts and the Coulomb gauge, we obtain
\[
\frac{\partial E_{\delta,n}}{\partial \delta}\Big|_{\delta=0} = - \frac{\hbar^2}{2m} \int_{-L}^L\mathrm{d}y~|\partial_x\psi_{0,n}(L,y)|^2\;.
\] 
Now we can easily generalize this result to the case when 
all the four walls are shifted outwards by $\delta$, and we have 
\begin{equation}
\frac{\partial E_{\delta,n}}{\partial \delta}\Big|_{\delta=0} = - \frac{\hbar^2}{2m} \oint_{\mathrm{wall}} |\boldsymbol{n}(s)\boldsymbol{\nabla}\psi_{0,n}(s)|^2\mathrm{d}s\;,
\label{pert:eq}
\end{equation}
where $s$ is an arc length parameter along the wall and $\boldsymbol{n}(s)$ is the corresponding normal vector. 
Therefore,~\eref{pert:eq} expresses the sensitivity of the energy levels on the side length of the box, and provides a useful method for selecting the magnetically localized states.

However, to estimate the change of the energy levels,  
this method can be applied only for small variations of wall size. 
Thus as a checkup, it is practical to compare the energy levels in case of two largely different wall sizes. 
Then, the obtained energy levels are regarded as magnetically localized states when they coincided approximately.

\section*{References}

\bibliographystyle{iopart-num}

\providecommand{\newblock}{}




\end{document}